\documentstyle[seceq,epsf,wrapft]{ptptex}

\title{$\Xi^-$ emission probabilities at $(K^-,K^+)$ reaction points \\
and the $\Xi N$ interaction}
\author{Yasuo {\sc Yamamoto}, Toru {\sc Tamagawa}$^{*}$
\footnote{Present address: RIKEN, Cosmic Radiation Lab.,
2-1 Hirosawa , Wako 351-0198, Japan}, 
Tomokazu {\sc Fukuda}$^{**}$ \\
and Toshio  {\sc Motoba}$^{**}$
}

\inst{
Physics Section, Tsuru University, Tsuru 402-8555, Japan
\\
$^*$Department of Physics, University of Tokyo, Tokyo 112-0033, Japan
\\
$^{**}$Laboratory of Physics, Osaka Electro-Communication University, \\
Neyagawa 572-8530, Japan
}

\recdate{
}

\abst{
$\Xi^-$ escaping and scattering probabilities in the $(K^-,K^+)$ reaction with
nuclear targets is investigated on the basis of the eikonal approximation.
Calculations are performed using the $S=-2$ sector
of the Nijmegen model-D interaction, whose hard core radii are
treated phenomenologically.
The $\Xi^- N$ elastic cross sections are 
derived from the $^9$Be$(K^-,K^+)$ data of the BNL-E906 experiment,
which is useful to determine the hard-core radii in $P$-states.
The $\Xi$ single-particle potential energy is investigated with the G-matrix 
calculation in nuclear matter, which is sensitive to the hard-core
radii in $S$-states. The important constraint for the $\Xi N$ 
interaction is obtained by combining the two analyses.
}

\begin{document}
\maketitle

\section{Introduction}
It is very important to study the properties of $\Xi N$ 
and $\Xi$-nucleus interactions in hypernuclear phenomena, 
because free-space $\Xi N$ scattering experiments are
difficult to perform with current experimental apparatus. 
Interesting information was obtained from 
events of simultaneous emissions of two
$\Lambda$ hypernuclei (twin $\Lambda$ hypernuclei)
in the KEK-E176 experiment.~\cite{Aoki93}  The obtained values of
binding energies $B_{\Xi^-}$ lead to an important conjecture for 
the $\Xi$-nucleus potential.
The BNL-E885 collaboration~\cite{E885} measured 
the missing mass spectra for the $^{12}$C$(K^-, K^+)$X reaction.  
Reasonable agreement between this data and theory is realized
by assuming a $\Xi$-nucleus potential $U_\Xi (\rho)=-V_0 f(r)$ with
well depth $V_0 \sim 14$ MeV within the Wood-Saxon (WS) prescription.

Recently, a new experiment (BNL-E906) has been performed in order to
detect charged particles emitted from $(K^-,K^+)$ reaction points
with a $^9$Be target using a cylindrical detector system.~\cite{E906}
Our concern in this work is with the emission probabilities of 
$\Xi^-$ particles produced through the quasi-free process in the target.
In recent analysis,~\cite{Tama}
$\Xi^-$ emission events have been separated into two categories,
the $\Xi^-$ escape process without any interaction after its production,
and the scattering process with a target nucleon, 
for which the obtained probabilities
are 63.6 $\pm$ 8.2\% (escaping probability) and 14.5 $\pm$ 2.6\% 
(scattering probability), respectively. From the scattering probability, 
the $\Xi^-N$ elastic scattering cross section in a nuclear medium was 
found to be 20.9 $\pm$ 4.5(stat)$^{+2.5}_{-2.4}$(syst) mb by using 
the eikonal approximation, in which the number of the
nucleons was eight and the reaction cross sections of $K^-$ and $K^+$
with the nucleons were 28.0 mb and 19.5 mb, respectively. 
Also, the elastic scattering probability was separated into the
$\Xi^- p$ part (5.8 $\pm$ 3.5\%) and the $\Xi^- n$ part (8.7 $\pm$ 4.4\%).

Nara et al. studied the $K^+$ momentum spectrum from $(K^-,K^+)$ reactions
in the framework of inter-nuclear cascade-type calculations.~\cite{Nara}
Here, two-step processes such as $K^-p \rightarrow M'Y$ followed 
by $M'N \rightarrow K^+ Y$, where $M'$ represents intermediate mesons,
play a significant role in $K^+$ yields, especially for heavy targets.
The contributions from these processes seem to be non-negligible, 
even in the high-momentum region of $K^+$, where quasi-free $\Xi^-$
productions are dominant. Because the $\Xi^-$ emission probabilities are
defined as the ratio of the number of emitted $\Xi^-$ particles to produced
$\Xi^-$ particles, the two-step contributions to $(K^-,K^+)$ events make 
them larger. The ratio of the two-step processes in the $(K^-,K^+)$ events,
denoted as $\alpha_{\rm two}$, is estimated to be about 10\% at most.~\cite{Ohnisi}
The above values of the escaping probability, 63.6\%, and the scattering
probability, 14.5\%, for $\alpha_{\rm two}=0$\% change to 67.1\% (71.0\%)
and 15.1\% (15.8\%), respectively, for $\alpha_{\rm two}=5$\% 
($\alpha_{\rm two}=10$\%).

In the report of the E906 experiment~\cite{E906}
the most important results of the theoretical calculations are 
cited,~\cite{YWFS92,YSW94,YWMF97} where a simple model
is proposed for the $\Xi^-$ emission process. 
The outline of this paper is as follows.
First, the plausibility of this model is investigated in light 
of the E906 data. Second, the observed $\Xi^-$ escaping and scattering 
probabilities are shown to give an important constraint 
for the $\Xi N$ interaction, which complements the $\Xi$ well depths 
derived from the data of E885 experiment.

\section{$\Xi^-$ emission probabilities and the $\Xi^- N$ elastic cross section}
We now formulate our model.~\cite{YWFS92,YSW94,YWMF97}
The transition rates $w_\Xi^{\rm scat}$ due to elastic scattering between 
a quasi-free $\Xi$ (momentum $\vec p_\Xi$) and nucleons 
(momenta $\vec p_N$) in nuclear medium are defined by
\begin{eqnarray}
w_\Xi^{\rm scat}&=& 
\int d^3 \vec p_N \int 
d^3 \vec p'_{\Xi N} \int d^3 \vec P'_{\Xi N}\ 
\,\rho \,f_N(\vec p_N)
\, v_{\Xi N}\,{d \sigma \over d \Omega} \cr 
&\times&
\delta ((p_{\Xi N}- p'_{\Xi N})/p'^2_{\Xi N})
\,\delta ^3(\vec P_{\Xi N}-\vec P'_{\Xi N})\,
\theta (p'_N-p_F) \theta (p'_\Xi-q_\Xi)  ,  
\label{eq:Wscat}
\end{eqnarray}
where $\rho$ is the nuclear density, ${d \sigma \over d \Omega}$ 
is the $\Xi N$ differential scattering cross section for the transition 
from the initial relative momentum $p_{\Xi N}$ to the final one 
$p'_{\Xi N}$, and $v_{\Xi N}$ is the relative velocity.  Also,
$\vec P_{\Xi N}$ and $\vec P'_{\Xi N}$ denote the initial and final
center-of-mass momenta, respectively. The function 
$f_N(p_N)=\big( {4 \over 3} \pi p_F^3 \big) ^{-1} \theta (p_F-p_N)$
represents a Fermi distribution of nucleons, with 
$p_F$ being the Fermi-momentum. 
The factor $\theta (p'_N-p_F)$ is the Pauli exclusion factor 
for the final nucleon state of momentum $p'_N$, and
$\theta (p'_\Xi-q_\Xi)$ allows transitions only to $\Xi$ states
whose momenta ($p'_\Xi$) are larger than the critical momentum $q_\Xi$ 
determined by ${q^2_\Xi \over 2M_\Xi}+U_\Xi(\rho) =0$, 
$U_\Xi(\rho)$ being a $\Xi$ single-particle potential in medium.
This constraint for $\Xi$ states results from the fact that 
lower-momentum components of scattered $\Xi$ particles are considered 
as being absorbed into target nuclei and not detected experimentally. 
In this work we assume $U_\Xi(\rho)= U_\Xi(\rho_0) (\rho/\rho_0)$, 
where $\rho_0$ is the normal density.
The value of $U_\Xi(\rho_0)$ is given roughly by the Wood-Saxon well depth 
of $\Xi$ in the E885 experiment.~\cite{E885} Here, we use 
$U_\Xi(\rho_0)=-14$ MeV,
but our results are not sensitive to the value of $U_\Xi(\rho_0)$.
Assuming isotropic angular distributions  
${d \sigma \over d \Omega}={\sigma^{el}_{\Xi N} \over 4\pi}$, 
Eq.~(\ref{eq:Wscat}) can be reduced to a compact form,~\cite{YWFS92}
where $\sigma^{\rm el}_{\Xi N}$ is the angle-integrated $\Xi N$ elastic
cross section. In the above expressions, charges of $\Xi$ particles and 
nucleons are implicit for simplicity. In actual calculations, we treat 
explicitly the processes $\Xi^- p \rightarrow \Xi^- p$ and 
$\Xi^- n \rightarrow \Xi^- n$. Then, the transition rates of $\Xi^-$ 
for protons and neutrons in the target are denoted as 
$w_{\Xi^-}^{\rm scat}(p)$ and $w_{\Xi^-}^{\rm scat}(n)$, and we have 
$w_{\Xi^-}^{\rm scat}=w_{\Xi^-}^{\rm scat}(p)+w_{\Xi^-}^{\rm scat}(n)$.

We define here the $\Xi^-$ absorption rate $w^{\rm abs}_{\Xi^-}$ phenomenologically.
This is determined from the probability of $\Xi^-$ vanishing  due to
processes such as $\Xi^- p \rightarrow \Xi^0 n$ and $\Xi^- p \rightarrow
\Lambda \Lambda$ in target nuclei. The corresponding $\Xi^-$ absorption 
cross section is denoted as $\sigma_{abs}$.
The $\Xi^-$ mean free path (MFP) $\lambda_{\Xi^-}$ in medium is given as follows.
The MFPs for scattering and absorption cross sections are represented
by $\lambda_{\Xi^-}^{\rm scat}=v_{\Xi^-}/w_{\Xi^-}^{\rm scat}$
and $\lambda_{\Xi^-}^{\rm abs}=1/(\rho \sigma_{\rm abs})$, where
$v_{\Xi^-}$ is the velocity of $\Xi^-$.
Then we have the relation 
$1/\lambda_{\Xi^-}=1/\lambda_{\Xi^-}^{\rm scat}+ 1/\lambda_{\Xi^-}^{\rm abs}$.
Using Eq.~(\ref{eq:Wscat}), $\lambda_{\Xi^-}^{\rm scat}$ is obtained
as a function of $p_{\Xi^-}$ and $\rho$. 

The probability of $\Xi^-$ reaction processes in a finite nucleus 
can be estimated on the basis of the eikonal approximation.
Assuming forward scattering, $\theta_{K^+}= \theta_{\Xi^-} = 0^\circ$,
the $(K^-,K^+)$ effective number for the reaction (scattering and
absorption) of the produced quasi-free $\Xi^-$ and nucleons is given by
\begin{eqnarray}
N_{\Xi^-}^{\rm reac}&=& 
\int_0^{\infty} 2\pi b\, db \int_{-\infty}^{\infty} dz\,
\rho (\sqrt{b^2+z^2})  \cr
&\times&
\exp \biggl\{-\bar \sigma_{K^- N}\int_{-\infty}^z \rho(\sqrt{b^2+z'^2})
 dz' -\bar \sigma_{K^+ N}\int_z^{\infty} \rho(\sqrt{b^2+z'^2})
 dz'\biggr\} \cr
&\times& \biggl\{1-\exp \Bigl(-\int_z^{\infty} {1 \over \lambda_{\Xi^-}}
 dz' \Bigr)
\biggr\} \ ,
\label{eq:Nreac}
\end{eqnarray}
with $1/\lambda_{\Xi^-}=1/\lambda_{\Xi^-}^{\rm scat}+ 1/\lambda_{\Xi^-}^{\rm abs}$.
Here, $\sigma_{K^- N}$ and $\sigma_{K^+ N}$ are the isospin averaged
total cross sections between $K^-$ and $K^+$ and a nucleon.
We set $\sigma_{K^- N}=28.0$ mb and $\sigma_{K^+ N}=19.5$ mb in this work.
The total effective number $N_{\rm total}$ for the $(K^-,K^+)$ reaction
is given by omitting the factor
$\Bigl\{1-\exp \Bigl(-\int_z^{\infty} {1 \over \lambda_{\Xi^-}} dz' \Bigr)
\Bigr\}$ in the above expression, and the reaction probability of $\Xi^-$ 
in medium is given by 
\begin{eqnarray}
P_{\Xi^-}^{\rm reac}=N_{\Xi^-}^{\rm reac}/N_{\rm total}  \ .
\end{eqnarray}
Similarly, we obtain the effective number $N_{\Xi^-}^{\rm abs}$ for
$\Xi^-$ absorption in medium by replacing 
$\lambda_{\Xi^-}$ in Eq.~(\ref{eq:Nreac}) with $\lambda_{\Xi^-}^{\rm abs}$.
Then, the scattering part of the effective number,
which originates from the $\Xi^- N$ scattering in medium,
is given as 
\begin{eqnarray}
N_{\Xi^-}^{\rm scat}&=& 
\int_0^{\infty} 2\pi b\, db \int_{-\infty}^{\infty} dz\, 
\rho (\sqrt{b^2+z^2})  \cr
&\times&
\exp \biggl\{-\bar \sigma_{K^- N}\int_{-\infty}^z \rho(\sqrt{b^2+z'^2})
 dz' -\bar \sigma_{K^+ N}\int_z^{\infty} \rho(\sqrt{b^2+z'^2})
 dz'\biggr\} \cr
&\times& \exp \Bigl(-\int_z^{\infty} {1 \over \lambda_{\Xi^-}^{\rm abs}}
 dz' \Bigr)
\biggl\{1-\exp \Bigl(-\int_z^{\infty} {1 \over \lambda_{\Xi^-}^{\rm scat}}
 dz' \Bigr)
\biggr\} \ .
\label{eq:Nscat}
\end{eqnarray}
Then the $\Xi^-$ scattering probability for the $\Xi^-$ emission
after $\Xi^- N$ scattering in medium is given by
\begin{eqnarray}
P_{\Xi^-}^{\rm scat}=N_{\Xi^-}^{\rm scat}/N_{\rm total}  \ .
\end{eqnarray}
The $\Xi^-$ escaping probability for $\Xi^-$ emission with no interaction 
after the production is given by
\begin{eqnarray}
P_{\Xi^-}^{\rm esc}=1- {N_{\Xi^-}^{\rm reac} / N_{\rm total}} \ .
\end{eqnarray}
The $\Xi^-$ emission probability is given as the sum of
the above two probabilities:
\begin{eqnarray}
P_{\Xi^-}^{\rm emit}&=&1-P_{\Xi^-}^{\rm abs}=P_{\Xi^-}^{\rm esc}+P_{\Xi^-}^{\rm scat}  \ , 
\end{eqnarray}
with the $\Xi^-$ absorption probability
$P_{\Xi^-}^{\rm abs}={N_{\Xi^-}^{\rm abs} / N_{\rm total}}$.
The probabilities $P_{\Xi^-}^{\rm esc}$, $P_{\Xi^-}^{\rm scat}$, $P_{\Xi^-}^{\rm emit}$
and $P_{\Xi^-}^{\rm abs}$ can be compared with the E906 data.
Here, it is very important in our treatment
that the experimental values of $P_{\Xi^-}^{\rm esc}$ and $P_{\Xi^-}^{\rm scat}$ 
be obtained separately in the analysis of the E906 data.
As another expression for the $\Xi^-$ scattering probability,
we define here the in-medium scattering cross section 
$\tilde \sigma_{\rm scat}$ by the relation 
$\lambda_{\Xi^-}^{\rm scat}=1/(\rho \tilde \sigma_{\rm scat})$.
Here, the value of $\tilde \sigma_{\rm scat}$ can be obtained
phenomenologically without using Eq.~(\ref{eq:Wscat}), by treating it
as a parameter in Eq.~(\ref{eq:Nscat}) to reproduce 
the data of $P_{\Xi^-}^{\rm scat}$.

The proton and neutron distributions in $^{9}$Be are represented as
$\rho(r)=\rho_0 (1+\alpha (r/1.65)^2) \exp (-(r/1.65)^2)$
with $\alpha$=1.46 (proton) and 2.10 (neutron)
on the basis of a theoretical calculation.~\cite{Okabe}

It is necessary to use appropreate $\Xi N$ interactions for calculations,
though at present we do not have sufficient knowledge of realistic models.
In this work we use mainly the $S=-2$ sectors of $SU(3)$-invariant 
OBE models, which have been used successfully to explore properties 
of $S=-2$ hypernuclear systems such as double-$\Lambda$ nuclei 
and $\Xi$-nucleus bound states.
Among the various interaction models, 
the Nijmegen D (ND)~\cite{NDF} seems to be suitable at present, 
because this model reproduce the attractive $\Xi^-$-nucleus potentials
found in the KEK-E885 experiment~\cite{E885} with a reasonable
choice of the hard-core radii $r_c$, as shown below.
ND should be considered here as a useful model for our purpose to
analyze the data of both the $\Xi^- N$ cross section
and the $\Xi$ well depth in medium complementarily.
Another $\Xi N$ interaction, which we use only for comparing with ND,
is of a purely phenomenological Wigner-type represented by 
a one-range Gaussian (ORG) form $v(r)=v_{\rm ORG} \exp (-(r/\beta)^2)$ 
with $\beta$ = 1.0 fm, for which the isospin dependence is not 
taken into account. 

Now our concern is in determining how the E906 data can be used to
determine hard-core radii $r_c$ in ND ($v_{\rm ORG}$ in ORG).
Hereafter hard-core radii in $S$-, $P$- and $D$-states are
denoted $r_c(S)$, $r_c(P)$ and $r_c(D)$.
The E906 data gives the $\Xi N$ scattering in medium 
at $\Xi$ momentum around 550 MeV/c, where the $P$-state
contributions are more important than the $S$-state contributions, 
and the results are sensitive to the value of $r_c(P)$, 
not only to that of $r_c(S)$.  Contrastingly, the $S$-state 
contributions are far more important than the $P$-state contributions
for $\Xi$ well depths in nuclei.
In the case of ND, therefore, the two data can be used, so that 
the hard-core radii of $S$- and $P$-states are determined separately.
First, the value of $r_c(S)=0.50$ fm is chosen so as to result in 
an appropriate attraction of the calculated $\Xi$ single-particle 
potentials, as shown in the following section.
Next, the value of $r_c(P)$ is determined so as to reproduce
the E906 data reasonably.
Additionally, we assume $r_c(D)=r_c(S)=0.50$ fm for simplicity.
The isospin dependence of the hard-core radii of ND is not taken
into account. It should be noted, however, that 
the isospin dependence in the OBE part of ND is treated precisely.
In the case that we use ND, the $\Xi N - \Lambda \Lambda$ coupling
interaction is treated exactly, but the couplings to $\Lambda \Sigma$
and $\Sigma \Sigma$ channels are ignored for simplicity. 

\begin{table}[t]
\begin{center}
\caption{In the cases of $\alpha_{\rm two}=$ 0.0, 0.05 and 0.10, the
values of $\sigma_{\rm abs}$ (mb) and $r_c(P)$ (fm) are chosen so as to 
reproduce the experimental values of $P_{\Xi^-}^{\rm abs}$, 
$P_{\Xi^-}^{\rm esc}$ and $P_{\Xi^-}^{\rm scat}$. The notation
$R(\Xi^- p/\Xi^- n)$ is used to represent 
$P_{\Xi^-}^{\rm scat}(p)/P_{\Xi^-}^{\rm scat}(n)$.
The values of $P_{\Xi^-}^{\rm esc}$ are $P_{\Xi^-}^{\rm scat}$ are 
the fitted ones and agree with the experimental values.
See the main text for the definitions of $\sigma_{\rm abs}$ (mb) and
$\tilde \sigma_{\rm scat}$ (mb).
}
\label{tab1}
\vskip 0.2cm
\begin{tabular}{|c|cc|c|cc|c|}\hline
$\alpha_{\rm two}$ & $\sigma_{\rm abs}$ & $\tilde \sigma_{\rm scat}$ &
$r_c(P)$ & $P_{\Xi^-}^{\rm esc}$(fitted) & $P_{\Xi^-}^{\rm scat}$(fitted) & 
$R(\Xi^- p/\Xi^- n)$  \\
\hline
0.00 & 21.1 & 21.6 & 0.480 & 0.636 & 0.145 & 0.706  \\
0.05 & 16.4 & 20.3 & 0.491 & 0.671 & 0.151 & 0.692  \\
0.10 & 11.7 & 19.1 & 0.502 & 0.709 & 0.158 & 0.680  \\
\hline
\end{tabular}
\end{center}
\end{table}

We now give our results.
There are essentially two free parameters in our formalism,
$\sigma_{\rm abs}$ and $r_c(P)$ in ND ($v_{\rm ORG}$ in ORG).
The former is chosen so as to reproduce 
the experimental value of $P_{\Xi^-}^{\rm abs}$, namely the probability
of vanishing of $\Xi^-$ in the $(K^-,K^+)$ reaction.
We have $P_{\Xi^-}^{\rm abs}=$ 0.219, 0.178, 0.133 
($\sigma_{\rm abs}=$ 21.1 mb, 16.4 mb, 11.7 mb)
for $\alpha_{\rm two}=$ 0.0, 0.05, 0.10, respectively.
Next $r_c(P)$ in ND ($v_{\rm ORG}$ in ORG) is determined 
so as to reproduce the experimental values of $P_{\Xi^-}^{\rm esc}$ and
$P_{\Xi^-}^{\rm scat}$. Table~\ref{tab1} gives the calculated values 
with ND at $p_{\Xi^-}=550$ MeV/c, around which the peak value of 
the $\Xi^-$ production strength is obtained.
Here, the experimental values of $P_{\Xi^-}^{\rm abs}$, $P_{\Xi^-}^{\rm esc}$
and $P_{\Xi^-}^{\rm scat}$ turn out to be reproduced precisely with the values
of $\sigma_{\rm abs}$ and $r_c(P)$ given in the table.
The in-medium scattering cross sections $\tilde \sigma_{\rm scat}$
are determined here so as to reproduce the data of $P_{\Xi^-}^{\rm scat}$
phenomenologically by using the relation 
$\lambda_{\Xi^-}^{\rm scat}=1/(\rho \tilde \sigma_{\rm scat})$,
instead of deriving $\lambda_{\Xi^-}^{\rm scat}$ from the $\Xi N$ interactions.

In the E906 experiment, the elastic scatterings between the produced 
$\Xi^-$ particles and nucleons in the target were separated into 
the $\Xi^- p$ part ($5.8\pm 3.5$\%) and the
$\Xi^- n$ part ($8.7\pm 4.4$\%).~\cite{Tama}  The ratio of the 
probabilities of $\Xi^- p$ and $\Xi^- n$ scatterings are regarded as
corresponding to 
$R(\Xi^- p/\Xi^- n)=P_{\Xi^-}^{\rm scat}(p)/P_{\Xi^-}^{\rm scat}(n)$
in our model. The calculated values are given in Table~\ref{tab1}.
These are found to be similar to the experimental value, $\sim 0.67$.
It is more reasonable, of course, to also consider the isospin 
dependence of $r_c(S)$ simultaneously. There is no information, 
however, allowing us to determine them independently at present.

Similar results are obtained with ORG. For instance, 
the above values of $P_{\Xi^-}^{\rm esc}$ and $P_{\Xi^-}^{\rm scat}$
for $\alpha_{\rm two}=0.0$ can be reproduced with $v_{\rm ORG}=-54.5$ MeV.
Because no isospin dependence is taken into account in ORG,
the obtained value of $R(\Xi^- p/\Xi^- n)=0.603$
is significantly different from that with ND.

\begin{figure}[th]
\epsfysize= 8cm
  \centerline{\epsfbox{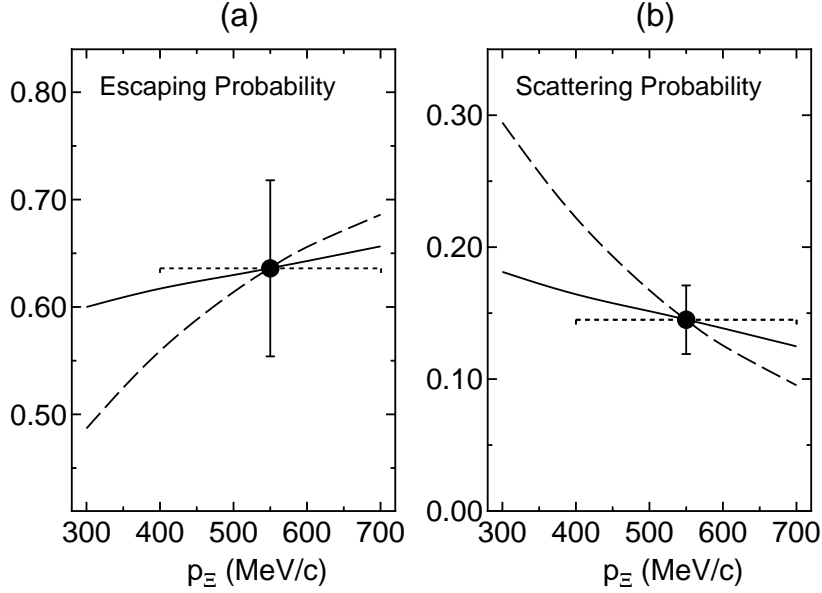}}
\caption{
Calculated values of $P_{\Xi^-}^{\rm esc}$ and $P_{\Xi^-}^{\rm scat}$ 
in the case $\alpha_{\rm two}$=0 as functions of $p_{\Xi^-}$ in (a) and (b).
Here, the solid and dashed curves correspond to ND and ORG.
The experimental values of $P_{\Xi^-}^{\rm esc}=63.6\pm 8.2$\% 
and $P_{\Xi^-}^{\rm scat}=14.5\pm 2.6$\% at $p_{\Xi^-}=550$ MeV/c
are shown for comparison, where the dotted horizontal bars 
indicate the width of the $\Xi^-$ spectrum (FWHM).}
\label{fig1}
\end{figure}

In Fig.~\ref{fig1}, the calculated values of the $\Xi^-$ escaping 
probabilities $P_{\Xi^-}^{\rm esc}$  and the $\Xi^-$ scattering probabilities
$P_{\Xi^-}^{\rm scat}$ are plotted as functions of $p_{\Xi^-}$ 
in (a) and (b), respectively.
Here, the solid and dashed curves correspond to ND and ORG.
The experimental values of $P_{\Xi^-}^{\rm esc}=63.6\pm 8.2$\% 
and $P_{\Xi^-}^{\rm scat}=14.5\pm 2.6$\% at $p_{\Xi^-}=550$ MeV/c
are shown for comparison, where the dotted horizontal bars 
indicate the width of the $\Xi^-$ spectrum (FWHM).
The values of $P_{\Xi^-}^{\rm esc}$ and $P_{\Xi^-}^{\rm scat}$ for ND and 
ORG turn out to be adjusted so as to reproduce the experimental values at 
$p_{\Xi^-}=550$ MeV/c, though their $p_{\Xi^-}$ dependences differ 
significantly. Thus, it is found that our two $\Xi N$ interactions
result in fits of equal quality by adjusting their interaction parameters.
As demonstrated in the next section, we need to complementarily analyze
the other data, such as those for the $\Xi$ well depths, in order to make
clear the features of the respective $\Xi N$ interactions.

\begin{figure}[th]
\epsfysize= 8cm
  \centerline{\epsfbox{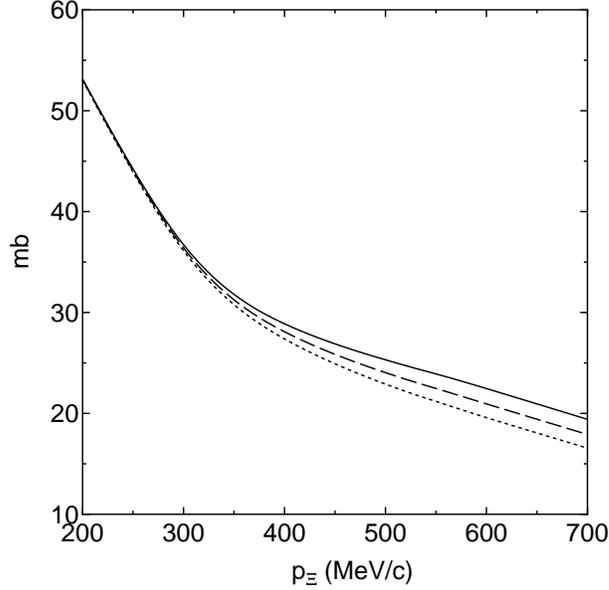}}
\caption{
Calculated values of $\Xi^- N$ elastic cross sections 
in mb are plotted as functions of the $\Xi^-$ momentum 
$p_{\Xi^-}$ (MeV/c) for ND with $r_c(P)=$ 0.480 fm, 0.491 fm
and 0.502 fm, by solid, dashed and dotted curves, respectively. 
The $\Xi^- N$ cross sections are obtained by averaging the $\Xi^- p$ 
and $\Xi^- n$ cross sections with the weight of 3 protons and 5 neutrons.
}
\label{fig2}
\end{figure}

In Fig.~\ref{fig2}, the calculated values of the $\Xi^- N$ elastic 
cross sections are plotted as functions of the $\Xi^-$ momentum 
$p_{\Xi^-}$ (MeV/c) for ND with $r_c(P)=$ 0.480 fm, 0.491 fm
and 0.502 fm, by solid, dashed and dotted curves, respectively.
The $\Xi^- N$ cross sections were obtained by averaging the $\Xi^- p$ 
and $\Xi^- n$ cross sections with the weight of 3 protons and 5 neutrons.
The ambiguity concerning the two-step processes is not serious
for the resulting $\Xi^- N$ cross sections.
This is essentially because $\Xi^- N$ scattering events were
observed explicitly in the E906 experiment, in spite of the
unknown two-step contributions.

There is an experimental error of 2.6\% for $P_{\Xi^-}^{\rm scat}=$ 14.5\%
in the case that we assume $\alpha_{\rm two}$=0; explicitly,
the upper and lower values of $P_{\Xi^-}^{\rm esc}$ are 17.1\% and 11.9\%.
Here, we attempt to reproduce the deviation of 2.6\% by changing 
only the value of $r_c(P)$: The values of 17.1\% and 11.9\% are 
reproduced by choosing $r_c(P)=$ 0.525 and 0.447, respectively. This 
choice is expressed as $r_c(P)=0.480^{+0.045}_{-0.033}$ fm for convenience.
In the above analysis, we assumed that the values of $\alpha_{\rm two}$ 
is between 0 and 10\%,  giving rise to $r_c(P)=$ 0.480 -- 0.502 fm.
Thus, it turns out that the ambiguity due to the two-step process
is less than the experimental error bar.

\section{$\Xi N$ G-matrix calculation and $U_\Xi$}
In order to demonstrate the properties of our interaction,
let us perform the $\Xi N$-$\Lambda \Lambda$ channel coupled 
G-matrix calculations in symmetric nuclear matter with ND 
whose hard-core radii are taken as the above values. Calculations
here are done in the framework used in a previous work~\cite{Yam94}:
We adopt the simple QTQ prescription for the intermediate-state
spectrum, which means that no potential term is taken into 
account in the off-shell propagation.

In the previous analysis with ND, $r_c(S)$ and $r_c(D)$ are
fixed at 0.5 fm, and $r_c(P)$ is adjusted so as to reproduce
the E906 data. Because our determined values of $r_c(P)$
are around 0.5 fm, it is reasonable here to use the simple
choice $r_c(S)=r_c(P)=r_c(D)=0.5$ fm. 
In this case, we calculate the single-particle potential energies 
$U_\Xi$ for a zero-momentum $\Xi$ at normal density.
The result is $U_\Xi=-14.3$ MeV, in which
the $S$-state, $P$-state and $D$-state contributions are
$U_\Xi(S)=2.8$ MeV, $U_\Xi(P)=-15.6$ MeV and $U_\Xi(D)=-1.6$ MeV,
respectively. It should be noted here that the negative
value of $U_\Xi$ is due to the strongly attractive 
$P$-state contribution.
As shown in the previous section, 
$P_{\Xi^-}^{\rm scat}= 14.5 \pm 2.6$\% corresponds to
$r_c(P)=0.480^{+0.045}_{-0.033}$, which
leads to $U_\Xi= -15.1^{-1.3}_{+1.8}$ MeV.
Thus, the ambiguity of $U_\Xi$ due to the experimental error of 
the E906 data is found to be small.

We now demonstrate that our obtained values of $U_\Xi$ are 
consistent with the experimental results. 
In the E885 experiment,~\cite{E885} the missing mass spectra
for $^{12}C(K^-, K^+)X$ reaction was measured.
Reasonable agreement between the data and theory is achieved 
by using a Wood-Saxon (WS) potential between $\Xi^-$ and the 
$^{11}$B core whose well depth is about 14 MeV.
In order to compare our result with this WS potential,
we construct the $\Xi$-nucleus potential from the calculated
values of $U_\Xi(\rho)$ with a local density approximation (LDA)
as 
\begin{eqnarray}
U_\Xi(r) &=& (t \sqrt{\pi})^{-3} \int d{\vec r}'
U_\Xi(\rho(r')) \exp (-|{\vec r} -{\vec r}'|^2/t^2)
\rho(r')  \ ,
 \label{xipot}
\end{eqnarray}
where the finite-range effect of $\Xi N$ interactions is 
taken into account by the parameter $t$.~\cite{Mah}
The nuclear density distributions $\rho (r)$ are obtained from
the Skyrme-Hartree-Fock calculations.
We compare our obtained $\Xi$-$^{11}$B potential with the above
WS potential by calculating the volume integral per nucleon,
$
J_V/A=-A^{-1} \int U(r) d^3 r ,
$
the root mean square radius,
$
\langle R_V^2 \rangle^{1/2} = 
\left[\int U(r) r^2 d^3r/ \int U(r) d^3r\right]^{1/2} ,
$
and the binding energy of the ground $1S$ state, $B_\Xi(1S)$,
obtained without taking the Coulomb interaction into account.

In the case of the above WS potential, the obtained values are
$J_V/A=133$ MeV fm$^3$, 
$\langle R_V^2 \rangle^{1/2}=3.1 $ fm and $B_\Xi(1S)=2.2 $ MeV.  
By setting $t=1.0$ fm in the case of our G-matrix interaction
derived from ND with $r_c(S)=r_c(P)=r_c(D)=0.5$ fm, we obtain 
$J_V/A=126$ MeV fm$^3$, $\langle R_V^2 \rangle^{1/2}=2.9 $ fm,
and $B_\Xi(1S)=2.2 $ MeV.
Thus, it is confirmed that the two $\Xi$-nucleus potentials are
very similar. This results from our choice of $r_c(S)=0.5$ fm,
because changes of $r_c(P)$ and $r_c(D)$ have 
far weaker effects on the $U_\Xi$ values.

The important information regarding the $\Xi$-nucleus potential
can be obtained from the events of simultaneous emissions
of two $\Lambda$ hypernuclei (twin $\Lambda$ hypernuclei)
in the KEK-E176 experiment.
These data give the energy difference between the initial
$\Xi^-$ bound state and the final twin $\Lambda$ state, 
namely the binding energy $B_{\Xi^-}$ between $\Xi^-$ and 
the nucleus.  
For instance, the event studied in Ref.~\citen{Aoki93}
indicates the $2P$ $\Xi^-$ bound state in $^{12}$C with 
$B_\Xi(2P)=0.57 \pm 0.19$ MeV. With the above procedure
we can derive the $\Xi^-$-$^{12}$C potential straightforwardly.
Including the Coulomb interaction, we obtain the calculated
value of 0.40 MeV for $B_\Xi(2P)$. This is slightly smaller
than the experimental value, but it is still within the error bar.

\begin{figure}[htbp]
  \epsfysize= 8 cm
  \centerline{\epsfbox{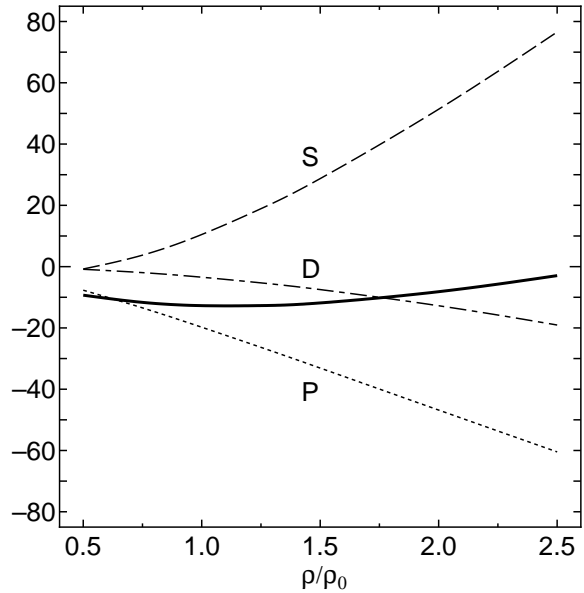}}
\caption{
The values of $U_{\Xi^-}$ in neutron matter
calculated with ND with $r_c(S)=r_c(P)=r_c(D)=0.5$ fm
are plotted as a function of $\rho/\rho_0$  by the solid curve.
$S$-, $P$- and $D$-state contributions are plotted by the
dashed, dotted and dot-dashed curves, respectively.
}
\label{fig3}
\end{figure}

Finally, let us point out the important contributions of the
$\Xi N$ $P$-state interaction to the single-particle potential 
energy of $\Xi$ in dense nuclear matter, which is
an interesting feature of ND.
Considering the important roles of negatively-charged baryons
in neutron stars, we calculate here the s.p.~potential
$U_{\Xi^-}$ in neutron matter.~\cite{YNT}
In Fig.~\ref{fig3}, the $U_{\Xi^-}$ values for 
ND with $r_c(S)=r_c(P)=r_c(D)=0.5$ fm are plotted
as a function of $\rho/\rho_0$  by the solid curve, where
$\rho_0=0.17$ fm$^{-3}$ is the normal density.
The $S$-, $P$- and $D$-state contributions are plotted by
the dashed, dotted and dot-dashed curves, respectively.
It should be noted here that the $S$-state contribution
is strongly repulsive and the negative values of $U_{\Xi^-}$
are due to the strongly attractive contribution of the
$P$-state interaction.
In the Ref.~\citen{YNT}, similar calculations were
performed using ND with hard-core radii assumed
to have the same values as the corresponding ones in
the $NN$ channel. The value of $r_c(P)=0.346$ fm in that case
is significantly smaller than the value $r_c(P) \sim 0.5$ fm
in the present case. This implies that the former $P$-state
interaction is too attractive.

\section{Conclusion}

The BNL-E906 experiment was performed in order to detect 
charged particles emitted from $(K^-,K^+)$ reaction points
on $^9$Be target. 
In recent analysis of the $\Xi^-$ emission events,
the $\Xi^-$ escape process without any interaction after 
its production and the scattering process with a target nucleon
have been extracted separately. 
These are specified by escaping and scattering probabilities,
respectively. A simple model was proposed 
for these processes,~\cite{YWFS92,YSW94,YWMF97}
which makes it possible to relate these probabilities with 
underlying $\Xi N$ interactions. 

Calculations were performed using the $S=-2$ sector of ND, 
where the hard-core radii are treated as free parameters.
The obtained $\Xi^-$ escaping and scattering probabilities 
are consistent with the E906 data,
where the choice of hard-core radii in the $P$-states is
sensitive to the results.

Another important piece of information is obtained from the
$\Xi$ single-particle potential energy, which is derived from  
the $\Xi N$ G-matrix calculations with ND in nuclear matter.
The resulting value is very sensitive to the hard-core
radii in the $S$-states.
Our obtained $\Xi$-nucleus potential, constructed with LDA, 
is similar to the WS potential obtained from the E885 experiment,

Thus, data from the E906 and E885 experiments
can be used complementarily, so that the hard-core radii 
of ND in $S$- and $P$-states can be determined separately.
The determined values of hard-core radii are 
not so different from those in $S=0$ and $S=-1$ channels.
This fact indicates that ND is a useful interaction model 
also in $S=-2$ channels, and we expect that it will 
be helpful to explore the effect of $\Xi^-$ mixing 
in neutron star matter.~\cite{Nishi}
Of course, there still remain some problems regarding modeling of ND.
Similar calculations are needed with use of more sophisticated
interaction models. The present study given will be 
an important test for such $\Xi N$ interaction models.

\section*{Acknowledgements}
This work was supported by a Grant-in-Aid for Scientific Research from
the Ministry of Education, Science, Sports and Culture (No. 11640286).

\end{document}